\documentclass[aps,prb,twocolumn,superscriptaddress]{revtex4-2}

\usepackage{amsmath,amssymb}
\usepackage{graphicx}
\usepackage{epstopdf}
\usepackage{bm}
\usepackage{xcolor}

\newcommand{\NTO}     {NiTe$_2$O$_5$}
\newcommand{\te}	{$^{125}$Te}
\newcommand{\slr} 	{$T_1^{-1}$}
\newcommand{\slrt} 	{$(T_1T)^{-1}$}
\newcommand{\hf} 	{$A_\text{hf}$}

\newcommand{\kk} 	{$\mathcal{K}$}

\newcommand{\TN} 	{$T_\text{N}$}

\begin{document}

\title{Persistence of Ising-like easy-axis spin correlations in the paramagnetic state of the spin-1 chain compound \NTO}

\author{Seung-Ho Baek}
\email[]{sbaek.fu@gmail.com}
%\homepage[]{Your web page}
\affiliation{Department of Physics, Changwon National University, Changwon 51139, Korea}
\affiliation{Department of Materials Convergence and System Engineering, Changwon National University, Changwon 51139, Korea}
\author{Jun Han Lee}
\affiliation{Department of Physics, Ulsan National Institute of Science and Technology, Ulsan 44919, Korea}
%\email[]{mycom3139@unist.ac.kr}
\author{Yoon Seok Oh}
%\email[]{ysoh@unist.ac.kr}
\affiliation{Department of Physics, Ulsan National Institute of Science and Technology, Ulsan 44919, Korea}
\author{Kwang-Yong Choi}
%\email[]{kchoi@cau.ac.kr}
\affiliation{Department of Physics, Sungkyunkwan University, Suwon 16419, Korea}
\author{Bernd B\"uchner}
%\email[]{b.buechner@ifw-dresden.de}
\affiliation{Institut f\"ur Festk\"orper- und Materialphysik and W\"urzburg-Dresden Cluster of Excellence ct.qmat, Technische Universit\"at Dresden, 01062 Dresden, Germany}
\affiliation{IFW Dresden, Helmholtzstr. 20, 01069 Dresden, Germany}

\date{\today}

%%%%%%%%%%%%%%%%%%%%%%%%%%%%%%%%%%%%%%%%%%%%%%%%%%%%%%%%%%%%%%%%%%%%

\begin{abstract}

A \te\ nuclear magnetic resonance (NMR) study was carried out in the paramagnetic state of the recently discovered quasi-one-dimensional spin-1 chain compound \NTO. We observed that the \te\ NMR spectrum  splits into two in a magnetic field applied along the $c$ axis. Based on the strong temperature variation of the relative intensity ratio of the split lines, we infer that the line splitting arises from the two sublattice susceptibilities induced in opposite directions along the chains. In great support of this interpretation, a quantitative analysis of the spin-lattice relaxation rate \slr\ and the Knight shift data unravels dominant transverse spin fluctuations. We conclude that Ising-like uniaxial spin correlations persist up to surprisingly high temperatures compared to the exchange energy scales. Spin-charge coupling mechanism via a self-doping effect may be important.

\end{abstract}

\maketitle

%%%%%%%%%%%%%%%%%%%%%%%%%%%%%%%%%%%%%%%%%%%%%%%%%%%%%%%%%%%%%%%%%%%%

\section{Introduction}

Weakly coupled one-dimensional (1D) spin chains  have long been a main subject of continuing interest as they provide a conspicuous ground for exploring emergent phenomena from many-body correlation effects \cite{steiner76, affleck89, vasiliev18}. 
A $S=\frac{1}{2}$ Heisenberg antiferromagnetic (AFM) chain with a nearest-neighbor (NN) intrachain interaction is well established to harbor a gapless disordered state \cite{steiner76} with fractional spinon excitations \cite{faddeev81,tennant93,mourigal13}. When there exist perturbations beyond an NN interaction, 
%induce a transition to long-range magnetic order at sufficiently low temperatures, 
exotic ground states such as frustrated spin liquids \cite{hase04,dutton12} and gapped spin singlet states \cite{hase93,seidel03,johnston87,azuma94} emerge, often displaying quantum critical behaviors \cite{lake05,blanc17,faure18} generated by strong 1D quantum fluctuations. % and peculiar spin excitations \cite{mikeska91,kim96} that have no analog in higher dimensions. 
For $S=1$ Heisenberg chains, on the other hand, spin correlations and spin excitations are quite distinct from those for $S=\frac{1}{2}$ ones in nature. 
For AFM chains, the Haldane gapped phase is stabilized \cite{buyers86,darriet93}, which entails a symmetry-protected topological state \cite{haldane83a}. In the ferromagnetic (FM) case, conventional magnon-like spin flip constitutes the elementary excitation %, of a FM $S=1$ chain, 
while quenching quantum fluctuations.  The typical examples include Y$_2$BaNiO$_5$ \cite{xu96}, Ni(C$_2$H$_8$N$_2$)$_2$NO$_2$(ClO$_4$) \cite{renard87}, and NiCl$_{2}$-4SC(NH$_2$)$_2$ \cite{zapf06} for the AFM case, as well as  NaCrGe$_2$O$_6$ \cite{nenert09} and NiNb$_2$O$_6$ \cite{chauhan20} for the FM one. The latter FM chains adopt a FM ground state, while
showing pronounced magnon-magnon interactions. However, little is known about dynamical spin excitations for FM $S=1$ chains with an AFM ground state due to the scarcity of candidate materials.

\begin{figure}
\centering
\includegraphics[width=\linewidth]{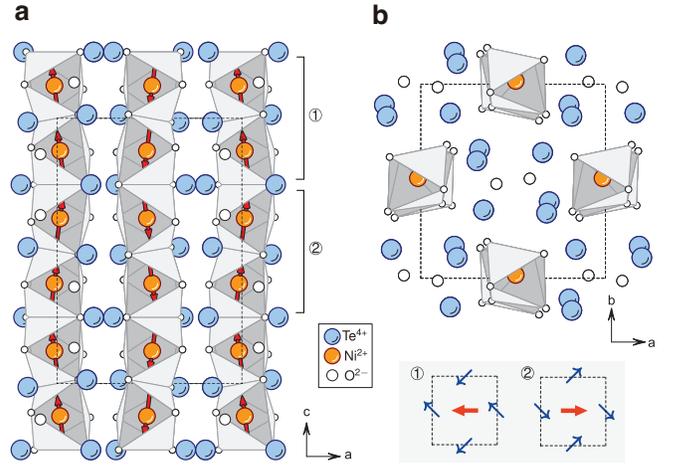}
\caption{Crystal structure of \NTO. (a) Side view along the $b$ axis with the spin configuration in the ordered state. (b) Top view along the chain direction. The tilting directions of the moments shown in (a) are $[\pm 1 {\pm1} 0]$ or $[\pm 1 {\mp1} 0]$ depending on the local distortion of the NiO$_6$ octahedra. The transverse components of the moments (blue thin arrows) generate a non-vanishing planar net moment along the $\pm a$ axis (red thick arrows), whose direction alternates every two layers. 
}
\label{structure}
\end{figure}

Recently, a new spin-1 quasi-1D chain system \NTO\ was reported \cite{lee19}. 
The crystal structure of \NTO\ (see Fig.~1) is orthorhombic (space group \textit{Pbnm}). The edge shared NiO$_6$ octahedra with Ni$^{2+}$ ($S=1$) form 1D chains oriented along the $c$ axis, which are well separated by lone-pair Te$^{4+}$ ions. 
A long-range AFM order takes place at $T_\text{N}\sim30.5$ K, in which the magnetic moments are ferromagnetically aligned parallel or antiparallel to the $c$ axis forming the ferromagnetic (FM) spin chains, while the NN chains are antiferromagnetically coupled as depicted in Fig.~1(a). 
We find that such an ordered spin structure is very rare in Heisenberg chain systems, being only occasionally realized in Ising-type chains with an easy axis anisotropy \cite{takeda71,niitaka01,coldea10}. %In comparison, a Heisenberg system with an FM intrachain coupling appears to adopt a FM ground state, e.g., NaCrGe$_2$O$_6$ \cite{nenert09} and NiNb$_2$O$_6$ \cite{chauhan20}.  
The FM coupled spin moments within the chain turn out to be weakly non-collinear, owing to the distortion of NiO$_6$ octahedra towards either $[\pm1 {\pm1} 0]$ or $[\pm1 {\mp1} 0]$ in which the sign change occurs every two octahedra. Interestingly, the resultant spin structure yields a non-vanishing net  planar moment along the $\pm a$ directions, as illustrated in the lower panel of Fig.~1(b). Consequently, an AFM planar moment sequence $\uparrow\uparrow\downarrow\downarrow\uparrow\uparrow\cdots$ is realized along the $c$ axis. % The non-vanishing planar moments may have a large influence on the low energy spin dynamics. 

In this paper, we examine the static and dynamic properties  of \NTO\ via \te\ NMR in its paramagnetic state. The Knight shift measurements reveal a strong anisotropy in the hyperfine coupling which supports the 1D character of the material. Strikingly, we observed the splitting of the \te\ line only for $H\parallel c$, which persists up to room temperature. Together with dominant transverse spin fluctuations deduced from the spin-lattice relaxation rate \slr, we conclude that the spin moments have a strong preference to point parallel or antiparallel to the chain direction retaining staggered AFM correlations up to temperatures much higher than the spin exchange energy. %We discuss that the Ising-like spin correlations with an easy axis anisotropy may arise from the self-doping effect of Ni$^{2+}$ ions. 

\section{Experimental details}

High-quality \NTO\ single crystals were synthesized using the flux growth method, as described in Ref. \cite{lee19}. The size of the crystal used in this NMR measurement is roughly $1\times 2\times 3$ mm$^3$. 

\te\ (nuclear spin $I=1/2$) NMR measurements were carried out at the external field of 10 T in the paramagnetic state (30--300 K). A one-axis goniometer was used to accurately align the sample to either the $c$ or [110] direction. 
The \te\ NMR spectra were obtained by a conventional Hahn spin-echo technique with a  typical $\pi/2$ pulse length of 3 $\mu$s. The nuclear spin-lattice relaxation time $T_1$ was measured by a saturation method, and extracted by fitting the nuclear magnetization $M(t)$ to a single exponential function, $1-M(t)/M(\infty)=A\exp(-t/T_1)$, where $A$ is a fitting parameter which is ideally unity. 

\section{Results}

\subsection{\te\ NMR spectrum and Knight shift}

\begin{figure*}
\centering
\includegraphics[width=0.8\linewidth]{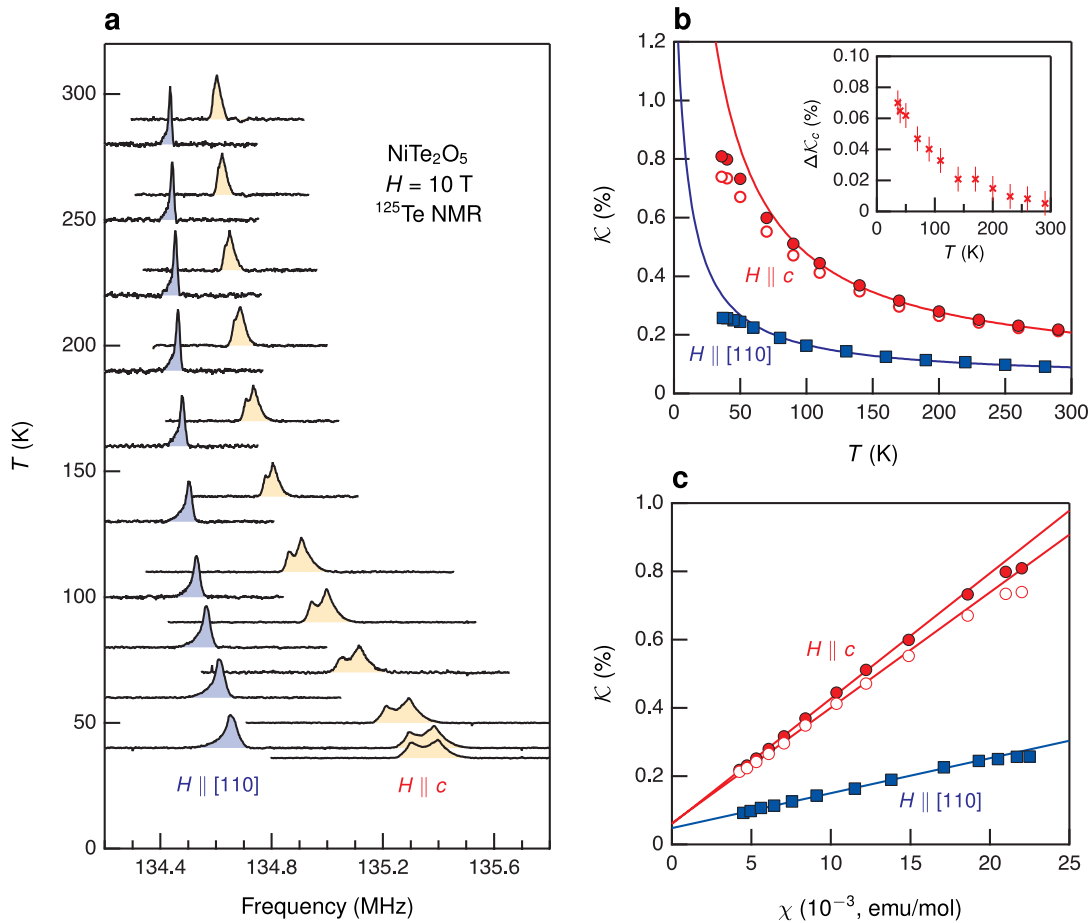}
\caption{(a) \te\ NMR spectra as a function of temperature measured at $\mu_0H=10$ T applied along [110] and [001], respectively. The line for $H\parallel c$ splits into two whose separation increases with decreasing temperature. The signal intensity of the lower peak may reflect a magnetization antiparallel to $H$ in the chain. 
(b) Knight shift \kk\ as a function of temperature and field direction. The closed and open circles represent the data for the upper and lower peaks, respectively. Solid lines are Curie-Weiss fits, which deviates below 100 K. Inset : temperature dependence of the difference in Knight shift for $H\parallel c$. 
(c) Knight shift \kk\ versus bulk uniform susceptibility $\chi$ plot, with temperature as the implicit parameter. The linear fits in the high-temperature region yield the hyperfine coupling constant \hf\ and the temperature-independent orbital shift $\mathcal{K}_0$ for the two field orientations (see text). 
}
\label{spec}
\end{figure*}

Figure 2(a) shows the temperature evolution of the \te\ NMR spectra measured at $\mu_0H=10$ T for two field orientations in the paramagnetic state of \NTO. While a moderate magnetic line broadening  toward \TN\ is found for both field directions [see Fig.~3(b)], the spectrum for $H \parallel c$ shows a much stronger temperature-dependent shift than that for $H\parallel [100]$ with decreasing temperature. Further, we find that the line for $H\parallel c$ splits into two at low temperatures, which sharply contrasts with the unsplit line for $H\parallel [110]$. 
For a quantitative analysis, the peaks of the spectra as a function of temperature are plotted in Fig. 2(b) in terms of the Knight shift \kk\ defined as $(\nu-\nu_L)/\nu_L\times 100$ \% where $\nu_L$ is the nuclear Larmor frequency. 
The solid lines are Curie-Weiss (CW) fits, i.e., $\mathcal{K}=\mathcal{K}_0 + C/(T+\theta)$ where $\theta$ is the CW temperature and $\mathcal{K}_0$ is the temperature-independent orbital shift. $\theta\sim 9$ K was determined from the extrapolated values of the linear fit of $1/\mathcal{K}$ vs.~$T$ in the high-temperature region (not shown), in excellent agreement with $\theta_\text{CW}=8.87$ K obtained by the bulk susceptibility analysis \cite{lee19}. 
The deviation from the CW behavior below $\sim 100$ K, as observed in the bulk susceptibility, is ascribed to the development of short-ranged AFM correlations.
It is noteworthy that $T_\text{N}/\theta \gg 1$ is extremely rare for a Heisenberg chain system and may be a signature of Ising-type spin correlations (single-ion anisotropy) \cite{steiner76} and an opposite sign of intrachain and interchain interactions.

Since the Knight shift is proportional to the local spin susceptibility $\chi_\text{loc}$, i.e., $\mathcal{K}=A_\text{hf}\chi_\text{loc}$ where \hf\ is the hyperfine coupling constant between the Te nuclei and the Ni$^{2+}$ moments, the strong anisotropy of the Knight shift, contrasting with the nearly isotropic bulk susceptibility $\chi$, is owing to the anisotropic \hf. 
In order to obtain \hf\ and $\mathcal{K}_0$, we plotted \kk\ vs.~$\chi$ with temperature as the implicit parameter in Fig. 2(c), assuming that the bulk susceptibility $\chi$ is equivalent to $\chi_\text{loc}$. 
The linear fit in the high-temperature region yields $A_\text{hf}=2.094$ and 1.873 kOe/$\mu_B$ for $H\parallel c$ and 0.57 kOe/$\mu_B$ for $H\parallel [110]$. In addition, the interception at $\chi=0$ of the linear fit gives the orbital shift $\mathcal{K}_0=0.062$ and 0.048 \%, respectively, which were applied in the CW-fit of \kk\ in Fig. 2(b). Since $\mathcal{K}_0$ is very small, the magnitude of the Knight shift almost entirely reflects the local spin susceptibility. 

%%%%%%%

\begin{figure}
\centering
\includegraphics[width=0.8\linewidth]{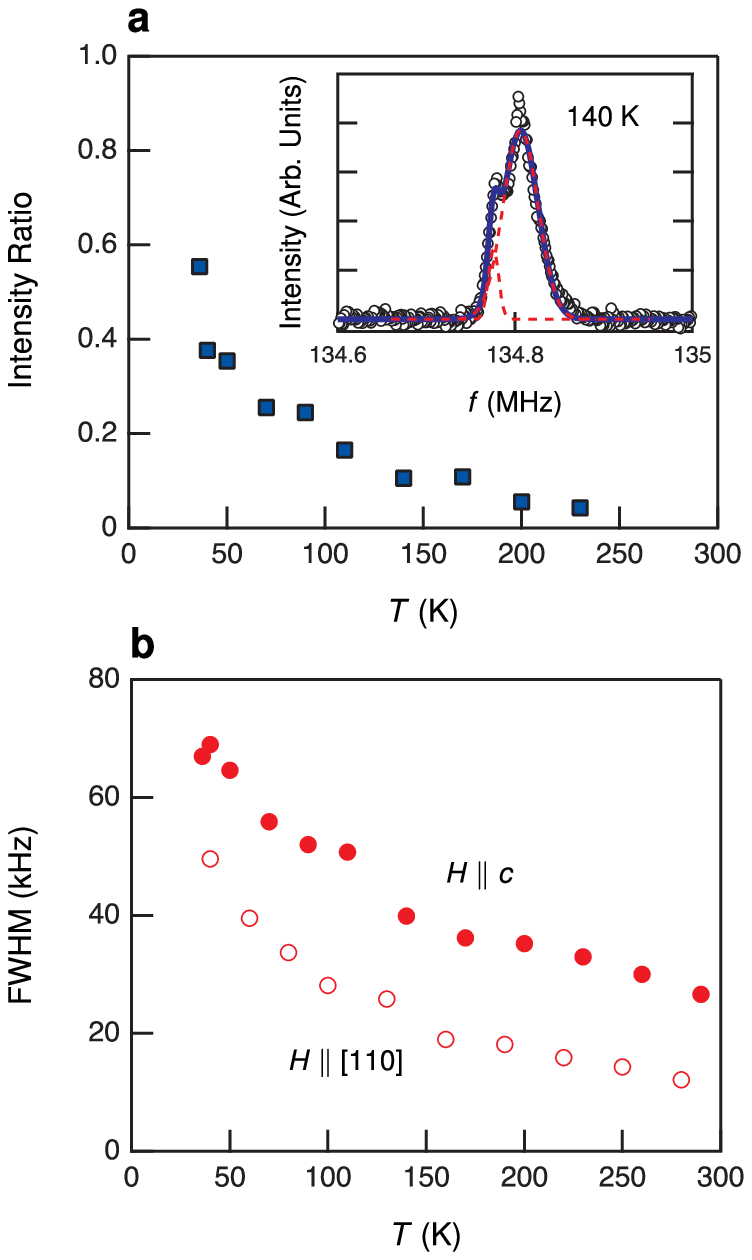}
\caption{(a) Intensity ratio between the two peaks observed for $H\parallel c$ as a function of temperature, obtained by the double Gaussian fit to the spectra. A fit to the 140 K spectrum is shown in the inset. 
(b) Temperature dependence of the full-width-at-half-maximum of the spectra for $H\parallel c$ (upper peak only) and $H\parallel [110]$.  It shows that the spectra for the two field orientations are similarly broadened with lowering temperature. 
}
\label{spec}
\end{figure}

The most remarkable feature seen in Fig. 2(a) is the clear NMR line splitting probed for $H\parallel c$. The amplitude of the splitting increases up to 90 kHz, or 6.7 mT in terms of hyperfine field strength. By contrast, the line for $H\perp c$ remains a single line down to \TN, which excludes the possible presence of two crystals as the origin of the splitting. Note that the asymmetric line shape with a shoulder for $H\perp c$ is essentially unchanged with temperature, while being magnetically broadened toward \TN\ as usual. Indeed, as shown in Fig. 3(b), the full-width at half maximum (FWHM) of the spectrum for $H\parallel [110]$ has the almost same temperature dependence as that for $H\parallel c$. 
At first sight, the line splitting for $H\parallel c$ may be ascribed to the presence of two inequivalent \te\ sites in a magnetic field due to different hyperfine couplings, assuming that magnetization is spatially uniform throughout the crystal. 
However, the strong temperature dependence of the relative intensity ratio between the two peaks [see Fig.~3(a)] brings this possibility into question. Clearly, the lower peak splits off from the upper one near room temperature and its relative intensity ratio rapidly grows with lowering temperature as approaching \TN. This observation is incompatible with the two \te\ sites with different \hf\ because the relative number of two nuclei, i.e., the relative signal intensity, should not change with temperature. Therefore, we conclude that the NMR splitting must be related to an intrinsic physical property which varies with temperature.
We stress that, since \te\ has the nuclear spin $I=1/2$ with no quadrupole moment, the splitting should be of magnetic origin, ruling out a possible role of the charge degree of freedom. Further, the temperature dependence of the line splitting for $H\parallel c$ in terms of the Knight shift difference [see the inset of Fig. 2(b)] reveals that the splitting follows a CW-like behavior as the Knight shift itself does, almost disappearing at room temperature. Such a Curie-type increase indicates that the line splitting does not represent a long-range order parameter, e.g. nematic order \cite{baek15}, but arises from a field-induced phenomenon.

Based on these observations, we reach a firm conclusion that the magnetic field applied along the $c$ direction develops two sublattice susceptibilities in chains, generating two \te\ sites which feel slightly different local fields. This implies that the staggered spin arrangement [see Fig.~1(a)] is strongly preferred in the normal state even up to room temperature. Namely, the upper (lower) peak may be linked to parallel (antiparallel) magnetization with respect to $H$.  
As the temperature is raised, thermal fluctuations would increase against the exchange interactions and thus tend to flip the antiparallel moments toward $H$, accounting for the rapid decrease of the relative intensity ratio between the split lines with increasing temperature [Fig. 3(a)]. 
%As the temperature is lowered, thermal fluctuations that tend to flip the antiparallel moments toward $H$ are progressively reduced, accounting for the rapid growth of the relative intensity ratio between the split lines toward \TN\ [Fig. 3(a)]. 
Furthermore, an additional AFM correlation among non-vanishing planar moments in the ordered state, shown in the lower panel of Fig.~1(b), may also persist in the normal state, seemingly accounting for the reduced bulk uniform susceptibility for $H\parallel a$ compared to that for $H\parallel b$ and $H\parallel c$ \cite{lee19}.

Note that, because the NMR line spitting arises from different local spin susceptibilities, the assumption of $\chi_\text{loc}=\chi$ made in Fig. 2(c) does not hold for $H\parallel c$, and thus it is reasonable to take the average of the two hyperfine couplings for $A_\text{hf}^{\parallel c}= 1.98$ kOe/$\mu_B$.

\subsection{Spin-lattice relaxation rate}

\begin{figure*}
\centering
\includegraphics[width=\linewidth]{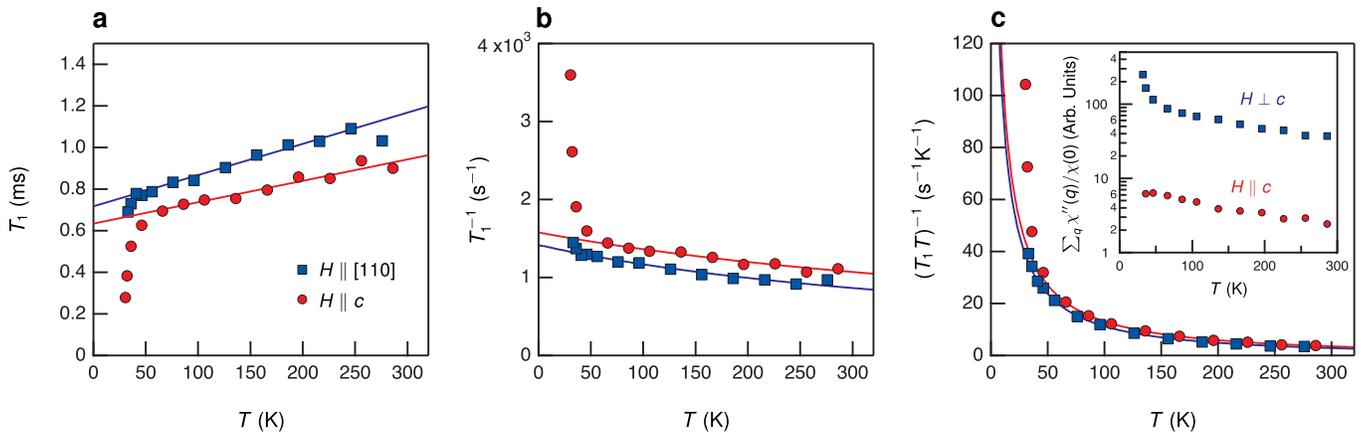}
\caption{(a) Temperature dependence of spin-lattice relaxation time $T_1$. The linear relation between $T_1$ and temperature down to $\sim 70$ K is observed for the two field orientations. (b) Spin-lattice relaxation rate \slr\ is described by $1/(T+T_\theta)$ where $T_\theta = 473$ and 630 K for $H\parallel c$ and $H\parallel [110]$, respectively. The critical slowing down of spin fluctuations causes the diverging behavior of \slr\ toward \TN\ below $\sim 70$ K. 
(c) \slrt\ is inversely quadratic in temperature. Inset: the ratio between the dynamical and uniform spin susceptibility $\sum_q\chi''(q)/\chi(0)$ as a function of temperature. It reveals that the dynamical susceptibility increases faster than the uniform one with decreasing temperature. The strong anisotropy of the dynamical susceptibility demonstrates the dominant planar spin fluctuations.}
\label{t1t}
\end{figure*}

Now we turn to the low-energy spin dynamics, which is probed by the spin-lattice relaxation rate \slr\ of \te. To begin with, we observed that $T_1$ is almost linearly decreased with decreasing temperature down to $\sim 70$ K. In other words, \slr\ above 70 K has a CW-like temperature dependence, $T_1^{-1} \propto 1/(T+T_\theta)$, where $T_\theta$ = 473 and 630 K for $H\parallel c$ and $H\parallel [110]$, respectively, were deduced from the linear fitting in Fig. 4(a). (Note that $T_\theta$ is totally irrelevant to the CW temperature $\theta$ deduced from the uniform susceptibility $\chi$ or the Knight shift \kk.) The sharp upturn of \slr\  below 70 K represents the critical slowing down of the spins towards the long-range AFM ordering at \TN. Note that the onset of the critical slowing down near $2T_N$ is typical in many magnetically ordered materials. 
Therefore, the weak enhancement of \slr\ at $T > 70$ K is not directly connected to the AFM ordering at \TN, but seems to arise from peculiar spin fluctuations of another energy scale. %We recall that the total magnetic entropy saturates for temperatures above 60 K \cite{lee19}. 

\slr\ divided by temperature, \slrt,  is a useful parameter because it is a measure of $\mathbf{q}$-averaged spin fluctuations perpendicular to the quantized axis. That is, \slrt\ can be written as,
\begin{equation}
(T_1T)^{-1}_{\parallel}  \propto \sum_\mathbf{q} [\gamma_N A_\text{hf}^{\perp} (\mathbf{q})]^2\chi_{\perp}''(\mathbf{q},\omega_L)/\omega_L, 	
\end{equation}
where $\gamma_N=13.454$ MHz/T is the gyromagnetic ratio of \te, $\chi''_\perp(q,\omega_L)$ is the imaginary part of the dynamical susceptibility at $\omega_L$, and the symbols $\parallel$ and $\perp$ represent the direction with respect to the quantization axis or the direction of the external field for a $I=1/2$ nucleus such as \te. 

The temperature dependence of \slrt\ is shown in Fig.~4(c). The fact that \slr\ has a $1/T$ dependence implies that the dynamical susceptibility $\sum_q\chi''(q)$ has an inversely quadratic temperature dependence, contrasting sharply with the uniform susceptibility $\chi(q=0)$ that is well described by the CW law. While \slrt\ probes only the dynamical susceptibility perpendicular to $H$, it is often possible to extract the component \textit{along the quantization axis} to estimate the anisotropy of spin fluctuations. 
For this purpose, we assume an axial symmetry about the $c$ axis and $q$-independence of $A_\text{hf}$. Then, one could approximate as $(T_1T)^{-1}_{\parallel c} = 2(A_\text{hf}^{\perp c})^2 \sum_q \chi_{\perp c}''(q)$ and $(T_1T)^{-1}_{\perp c} = (A_\text{hf}^{\parallel c})^2 \sum_q \chi_{\parallel c}''(q)+(A_\text{hf}^{\perp c})^2 \sum_q \chi_{\perp c}''(q)$, where the constant prefactor including $\gamma_N^2/\omega_L$ was set to one for simplicity. Relating them to the Knight shift $\mathcal{K}_\alpha =A_\text{hf}^\alpha \chi_\alpha(0)$ where $\alpha={\parallel c},{\perp c}$, and using the hyperfine coupling constants $A_\text{hf}^{\parallel c}=1.98$ and $A_\text{hf}^{\perp c}=0.57$ kOe/$\mu_B$ obtained above, one could estimate the ratio $\sum_q \chi''(q)/\chi(0)$  for the longitudinal ($H \parallel c$) and transverse ($H \perp c$) directions as a function of temperature. 

The results are presented in the inset of Fig. 4(c). The increase of $\sum_q \chi''(q)/\chi(0)$ with decreasing temperature is clearly seen for both directions, verifying the inverse quadratic temperature dependence of the dynamical susceptibility.
A remarkable observation is that $\sum_q \chi''(q)/\chi(0)$  is very anisotropic. Considering that $\chi(0)$ is nearly isotropic, this result implies that the transverse (planar) component of the dynamical susceptibility is an order of magnitude stronger than the longitudinal one in the whole temperature range investigated.% Furthermore, we find that the magnitude of the anisotropy is nearly unchanged up to room temperature.

\section{Discussion}

The NMR report about the anisotropy of low-energy spin fluctuations in the normal state well above $T_\text{N}$ is hardly found in the 1D chain systems and thus it is difficult to find a reference system to compare with. Regardless,
the dominant planar spin fluctuations over the longitudinal component suggest that the spin moments of Ni$^{2+}$ have a strong tendency to line up along the chain direction. 
Surprisingly, this result is well consistent with the presence of uniaxial sublattice susceptibilities inferred from the NMR line splitting for $H\parallel c$. Namely, both the static and dynamic NMR parameters suggest that spin correlations in \NTO\ are Ising-like with an easy-axis anisotropy involving the FM intrachain and AFM interchain couplings. 
The persistence of the Ising-like spin correlations up to room temperature ($\sim10T_\text{N}$) is striking, given the relatively small effective exchange $J \sim 6.63$ K \cite{lee19}, although the intrachain and interchain exchange couplings cannot be separately obtained from the CW fitting.
%At any rate,  we can impose $|J_\text{nnn}|<|J_\text{nn}|$ from the positive CW temperature and the single-ion anisotropy $D$ is usually not larger than $J_{\mathrm{nn}}$ for Ni$^{2+}$ ions. 
At any rate, the CW analysis of both the bulk susceptibility and the Knight shift shows that the spin exchange energy scale cannot exceed $\sim 100$ K at which short-range spin order sets in. This assertion is further supported by the fact that the magnetic entropy is released below 60~K \cite{lee19}.

Although the physical origin for the Ising-like anisotropy that is robust even at $10T_\text{N}$ is unclear, we speculate 
%Before proceeding, we recall that for $S=1/2$ chains, \slr\ is dictated by staggered and uniform spin fluctuations in the low- and high-temperature regimes, respectively. The uniform component gives rise to a linear increase of \slr\ with temperature \cite{sandvik95a,takigawa96a}. At high temperatures, our \slr\ data show a linear decrease, which is opposite to the linearly increasing behavior of the $S=1/2$ case. This demonstrates that the spin dynamics and dynamical fluctuations of \NTO\  are distinct from those of a $S=1/2$ AFM chain.
that the coupling between Ni$^{2+}$ ions and the ligand O$^{2-}$ ions could play an important role. The electron state of Ni$^{2+}$ is ideally a $d^8$ configuration, but the actual average valence per Ni ion may be slightly less than 2+ (two holes). This is equivalent to a self-hole-doping effect, with the partial occupation of the charge-transfer state $d^9\underbar{L}$ where $\underbar{L}$ denotes the ligand hole \cite{imada98}. The intermediate $d^9\underbar{L}$ state may be viewed as a local hole pair with the same spin, or the Zhang-Rice triplet ($S=1$) in analogy to the Zhang-Rice singlet picture in hole-doped cuprates \cite{zhang88}. 
%With a negligibly small or negative charge-transfer gap, the hole pair may easily hop to the nearest-neighbor site along the chain. Consequently, easy-axis single-ion anisotropy of the Ni$^{2+}$ ions,  which is otherwise small in an octahedral field, may be considerably strengthened by the spin-charge coupling. 
Remarkably, strong charge-spin coupling is observed in van der Waals antiferromagnet NiPS$_3$,
which shows a spin-orbit-entangled exciton stemming from the Zhang-Rice triplet \cite{kim18a,kang20}.
If \NTO\ has a self-doped ground state as NiPS$_3$, one can expect a close coupling
of electronic and magnetic structures. In this case, charge fluctuations are not completely
quenched. Rather, spin-charge-coupled fluctuations could affect a dynamical spin susceptibility
while preserving Ising-like correlations in the paramagnetic state.  

Another unusual finding is the inverse temperature dependence of \slr\ [Figs. 4(a) and (b)]. 
In the paramagnetic limit with localized moments, i.e., $T\gg J_\text{ex}$ where $J_\text{ex}$ is the effective exchange coupling, \slr\ is proportional to $A_\text{hf}^2/J_\text{ex}$ \cite{moriya56}. 
Since \hf\ is well-defined in a wide temperature range as shown in Fig. 2(c), one may conjecture that $J_\text{ex}$ decreases with decreasing temperature, causing the $1/T$ dependence of \slr. 
In this case, the reduction of $J_\text{ex}$ may be caused by the development of spin fluctuations at finite $q$ \cite{baek14a} and/or by the single-ion anisotropy enhanced by the spin-charge coupling mechanism conjectured above.
If the localized limit is not appropriate to describe the spin dynamics of \NTO\ in the temperature range investigated, however, the $1/T$-dependence of \slr\ may reflect the peculiar spin dynamics of \NTO. For comparison, we recall that for $S=1/2$ chains, \slr\ remains a constant in the quantum critical regime ($T \ll J_\text{ex}$) \cite{ishida96a,takigawa96a} being attributed to fermionic thermal excitations \cite{sachdev94,sandvik95a}. 
 One may note that the CW fitting of \slr\ in Fig. 4(b) yields $T_\theta\sim 500-600$ K, which appears to be the desired energy scale for the persistent uniaxial spin correlations through the spin-charge coupling at high temperatures.

%%%%%%%%%%%%%%%%%%%%%%%%%%%%%%%%%%%%%%%%%%%%%%%%%%%%%%%%%%%%%%%%%%%%

\section{Summary}

In conclusion, by means of \te\ NMR, we have investigated the static and dynamic spin susceptibilities in the paramagnetic state of the spin-1 chain compound, \NTO. Our measurements imply that easy-axis Ising-like spin correlations with FM intrachain and AFM interchain exchange couplings persist even at temperatures much higher than \TN. This finding is surprising, because it requires an additional intriguing interaction term with a much larger energy scale than the nearest-neighbor exchange coupling. The unusual inverse temperature dependence of \slr\ in the normal state may be an intrinsic dynamical feature of \NTO. 
Spin-charge coupling arising from a self-hole-doping mechanism may be an important term in the spin Hamiltonian. 
Clearly, future optical spectroscopic experiments in conjunction with theoretical
studies are called for to disclose an intertwining of electronic and magnetic degrees of freedom and to understand the unusual spin correlations in \NTO. 

%%%%%%%%%%%%%%%%%%%%%%%%%%%%%%%%%%%%%%%%%%%%%%%%%%%%%%%%%%%%%%%%%%%%

\begin{acknowledgments}
This work was supported by the National Research Foundation of Korea (NRF) grant funded by the Korea government(MSIT) (Grant No. NRF-2020R1A2C1003817).
\end{acknowledgments}

%%%%%%%%%%%%%%%%%%%%%%%%%%%%%%%%%%%%%%%%%%%%%%%%%%%%%%%%%%%%%%%%%%%%

\bibliography{mybib}

\end{document}